\documentclass{webofc}
\usepackage[varg]{txfonts}   

\begin{document}

\title{HI and H$_2$ gas evolution over cosmic times: {\sc ColdSIM} } 

\author{\firstname{Umberto} \lastname{Maio}\inst{1,2}\fnsep\thanks{\email{umberto.maio@inaf.it}} }

\institute{INAF, Italian National Institute of Astrophysics, via G. Tiepolo 11, 34143 Trieste, Italy
\and
	Max Planck Institute for Astrophysics, Karl-Schwarzschild-Str. 1, 85748 Garching b. M., Germany
          }
\abstract{
We present first results of cold cosmic gas evolution obtained through a set of state-of-the-art numerical simulations ({\sc ColdSIM}).
We model time-dependent atomic and molecular non-equilibrium chemistry coupled to hydrodynamics, star formation, feedback effects, various UV backgrounds as suggested by the recent literature, HI and H$_2$ self-shielding, H$_2$ dust grain catalysis, photoelectric heating and cosmic-ray heating.
By means of such non-equilibrium calculations we are finally able to reproduce the latest HI and H$_2$ observational data.
Consistently with available determinations, neutral-gas mass density parameter results around $ \Omega_{\rm neutral} \sim  10^{-3}$ and increases from lower to higher redshift ($z$).
The molecular-gas mass density parameter shows peak values of $ \Omega_{\rm H_2} \sim 10^{-4}$, while expected H$_2$ fractions can be as high as $ 50\% $ of the cold gas mass at $z\sim 4$-8, in line with the latest measurements from high-$z$ galaxies.
These values agree with observations up to $z\sim 7$ and both HI and H$_2$ trends are well reproduced by our non-equilibrium H$_2$-based star formation modelling.
Corresponding H$_2$ depletion times remain below the Hubble time and comparable to the dynamical time at all epochs. This implies that non-equilibrium molecular cooling is efficient at driving cold-gas collapse in a variety of environments and since the first half Gyr.
Our findings suggest that, besides HI, non-equilibrium H$_2$ analyses are key probes for assessing cold gas and the role of UV background radiation.
}
\maketitle
\section{Introduction}
\label{intro}

Atomic and molecular gas are fundamental phases of the cosmic medium since they constitute the bulk of neutral cosmic gas and lead to star formation and cosmic structure evolution.
Neutral gas is constituted by gas with temperatures below $10^4 \, \rm K$ (i.e. cold gas), while its chemical composition is characterised by neutral H atoms (HI), as well as by large amounts of H$_2$ molecules, the most abundant ones in the Universe.
These latter drive gas collapse and its conversion into stars.
Feedback effects from newly born structures, then, act on existing gas via a number of additional processes, such as supernova (SN) explosions, winds, metal enrichment, UV photoionization or photodissociation.
These can affect HI and H$_2$ evolution dramatically.
Observationally, HI is currently well determined up to high redshift, $z\simeq 5$, instead H$_2$ is being constrained up to $z\simeq 7$ by the latest determination based on e.g. ALMA, VLA, NOEMA, UKIRT, etc..
From a theoretical point of view, chemistry abundances in the cosmic space evolve off equilibrium conditions, over an expanding background, and for this reason species number densities must be computed considering explicitly the related time variations (non-equilibrium chemistry).
HI is expected to dominate cold material, because recombination processes are efficient around or below $10^4\,\rm K$ in UV shielded gas.
H$_2$ is more troublesome to study, since its formation in the cosmic space takes place through several channels.
In presence of free electrons and protons, H$_2$ can be formed via H$^-$ and H$_2^+$ catalysis.
In dense cold neutral gas, three-H processes are also effective for H$_2$ production.
After metal spreading during stellar evolution and SN events, heavy elements can condensate into dust grains and enhance H$_2$ abundances.
H$_2$ dust grain catalysis is metal-dependent and usually accompanied by photoelectric heating.
Besides metals, SNe can produce cosmic rays, too, and these further influence H$_2$ evolution.
Albeit studied and observed in the local Universe, these processes are poorly known in cosmological environments and deserve careful investigations.
Moreover, the establishment of a cosmological UV background at different $z$ can alter chemical species in a way that is not trivial to predict.
In the following, we outline the physical and chemical modelings required to address cosmic atomic and molecular gas in the most generic non-equilibrium conditions (Sect.~\ref{sec:method}).
We discuss the main results obtained for gas mass density parameters according to different assumptions and their comparisons to observational HI and H$_2$ determinations (Sect.~\ref{sec:results}). Then, we draw our conclusions (Sect.~\ref{sec:conclusions}).
We adopt a flat $\rm\Lambda$CDM cosmological model with present-day 
expansion parameter normalised to 100~$\rm km/s/Mpc$ of  
$h=0.7$,  
and baryon, matter and cosmological-constant density parameters of  
$\Omega_{\rm 0,b}=0.045$, 
$\Omega_{\rm 0,m}=0.27$    
and  $\Omega_{\rm 0,\Lambda}=0.73$,     
respectively.
The $z=0$ cosmological critical density is
$\rho_{0,crit} \simeq 277.4 \, h^2\, \rm M_\odot/kpc^3$.

\section{Method}
\label{sec:method}
Numerical treatment for cosmic gas and structure formation is implemented in an updated version of the P-Gadget3 code (based on \cite{Springel2005}) that solves the equations for gravity and smoothed particles hydrodynamics.
Non-equilibrium number densities are followed for e$^-$, H, H$^+$, H$^-$, He, He$^+$, He$^{++}$, H$_2$, H$^+_2$, D, D$^+$, HD, HeH$^+$ by solving first-order differential equations that  provide the temporal variations of each species, according to the relevant creation and destruction processes \cite{Maio2007, Maio2010, Maio2011, Maio2015}.
Besides atomic ionizations and recombinations, H$_2$ formation is followed according to the 
H$^-$ and H$_2^+$ channels, as well as three-body processes.
To account for density-dependent HI self-shielding we fit available tabulated data \cite{Rahmati2013}, while H$_2$ self-shielding is based on \cite{DB1996}, but we also check that more recent shielding formulations lead to similar results.
The role of different UV backgrounds at different epochs is assessed employing photoheating and photoionisation rates suggested by 
\cite{HM1996} (HM), \cite{P2019} (P19) and \cite{FG2020} (FG20).
These implementations are needed to follow molecule evolution during cosmic gas cooling and star formation in pristine regimes.
Star formation in dense shielded regions, stellar feedback and heavy-element production (C, N, O, Ne, Mg, Si, S, Ca, Fe, etc.) from stars with different masses and metallicities during SNII, AGB and SNIa phases is accounted for, as well \cite{Tornatore2007, Maio2010}.
To include H$_2$ evolution in enriched environments, we additionally consider chemical rates and energy transfer of dust grain catalysis and photoelectric heating 
\cite{HM1979, BT1994, Cazaux2004, Omukai2005}.
Dust grain temperature, $T_{\rm gr}(z) $, is estimated assuming a power-law emission at each $z$ \cite{DL1984}, although adopting either a constant value in the range $T_{\rm gr} = 40$-120~K or $z$-dependent values equal to the CMB temperature
-- i.e., $T_{\rm gr}(z) = T_{\rm CMB} (z)$ -- 
does not affect significantly the final results.
Dust chemistry and cooling/heating rates scale linearly with gas metallicity, as expected from local-Universe studies.
Cosmic-ray heating in star forming regions is modelled following \cite{Padovani2018} and scaling the transferred energy by the local star formation rate.
By means of such modelings, we run cosmological boxes of 10~Mpc/$h$ side length in which the initial gas and dark-matter density fields are sampled by 512$^3$ particles for each species.
Stellar particles are spawned when gas densities reach at least 10~$\rm cm^{-3}$ and the resulting stellar evolution is based on a reference Salpeter initial mass function (IMF), although we checked that adopting an alternative Chabrier IMF results are affected by less than a factor of 2.
SN feedback takes place with a reference efficiency of 0.1, while wind feedback is modelled by a constant wind velocity of 350~km/s. Spreading of heavy elements is mimicked by smoothing the above-mentioned individual metallicities over the hydrodynamic kernel.
Let us note that for a detailed picture one should be able to resolve both rare structures on large scales and small objects on tiny scales.
Because of numerical feasibility, a fair trade-off to assess non-equilibrium chemistry, halo statistics and luminosity functions is often achieved by employing boxes of the kind considered here.
Larger boxes could be used to track better massive rare structures, but this would come at detriment of gas chemistry features, as also demonstrated in the literature.
Reducing the box size to obtain better resolution would be a viable option to follow more precisely atomic and molecular species, but, besides statistical issues, this would imply very long computing times.
With our set-up we can already reach H$_2$ mass fractions of $ \sim 50\% $ in dense collapsing regions, thus any issue related to box size or resolution would impact our results by less than a factor of 2.
This is better than the uncertainties on particular chemical rates, IMF and physical processes at low metallicities (e.g. gas grain processes).
Unknowns in stellar evolution, such as exact yield determinations, stellar rotation, impacts of the dredge-ups, AGB mass loss, etc. can have some effects. Here, we follow all the relevant stellar phases consistently with mass-dependent yields and lifetimes, hence our findings should be robust.

\section{Results}
\label{sec:results}
\begin{figure*} 
\centering
\includegraphics[scale=0.32, trim=0 15 0 30, clip]{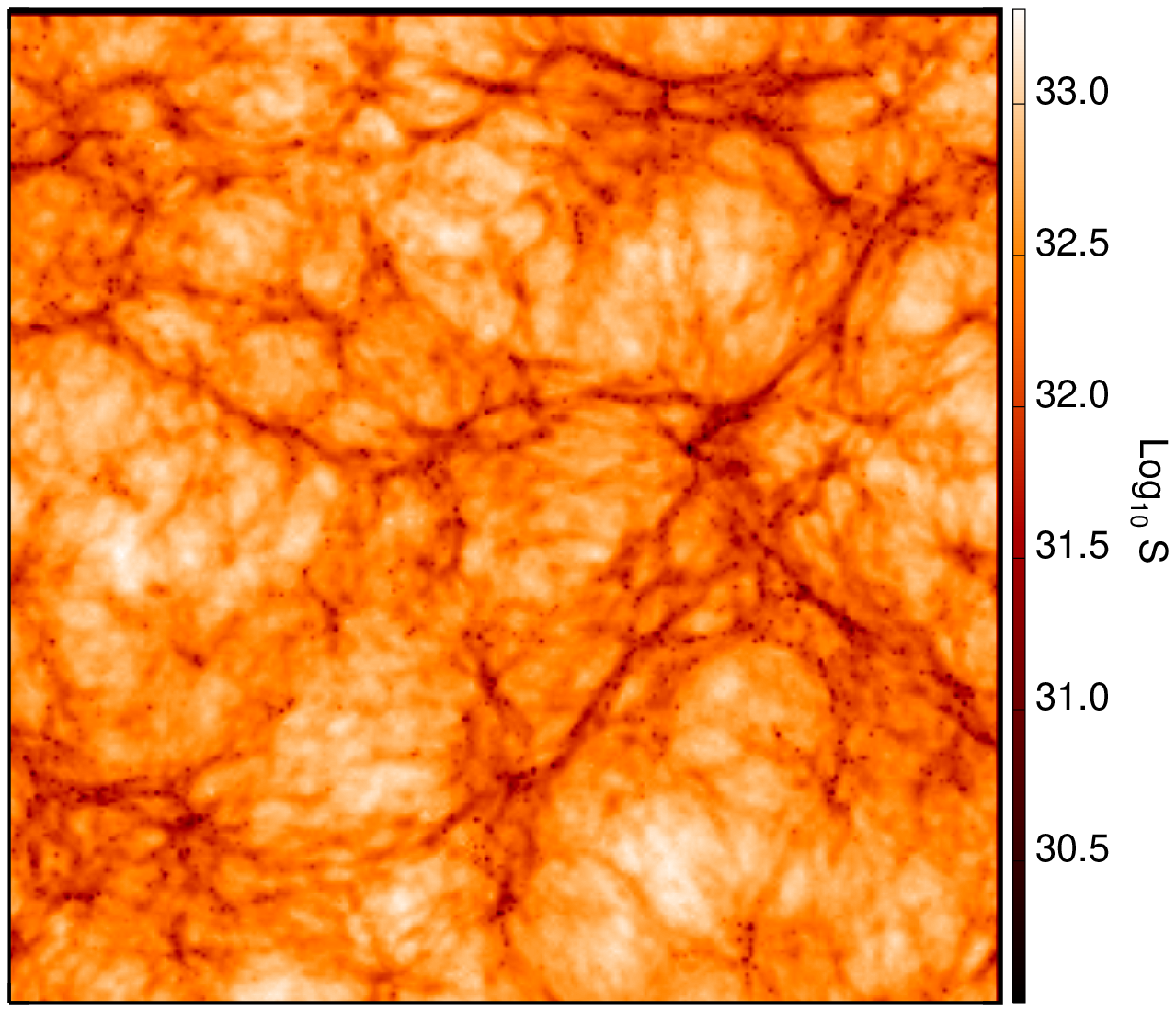}
\includegraphics[scale=0.32, trim=0 15 0 30, clip]{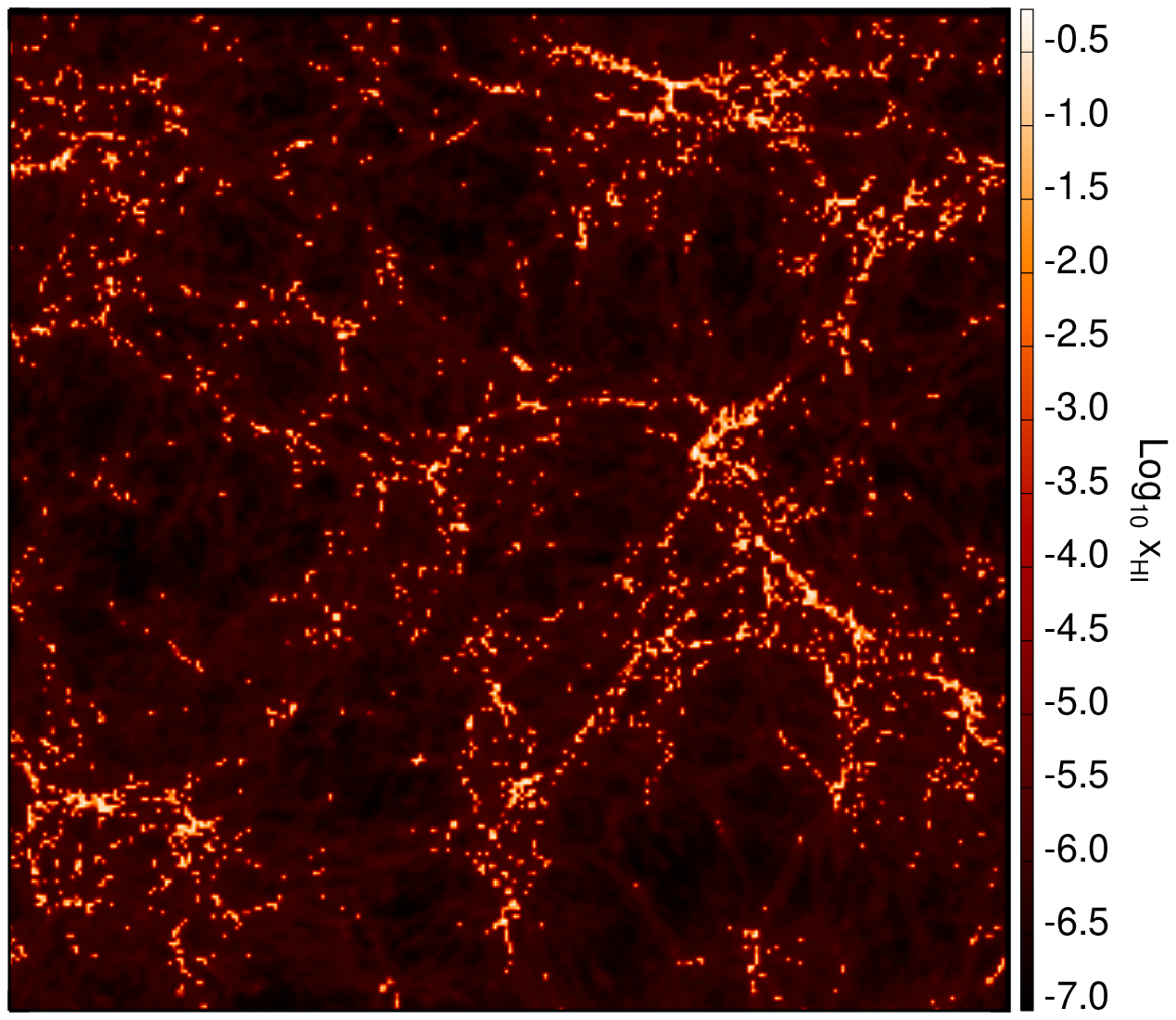}
\includegraphics[scale=0.32, trim=0 15 0 30, clip]{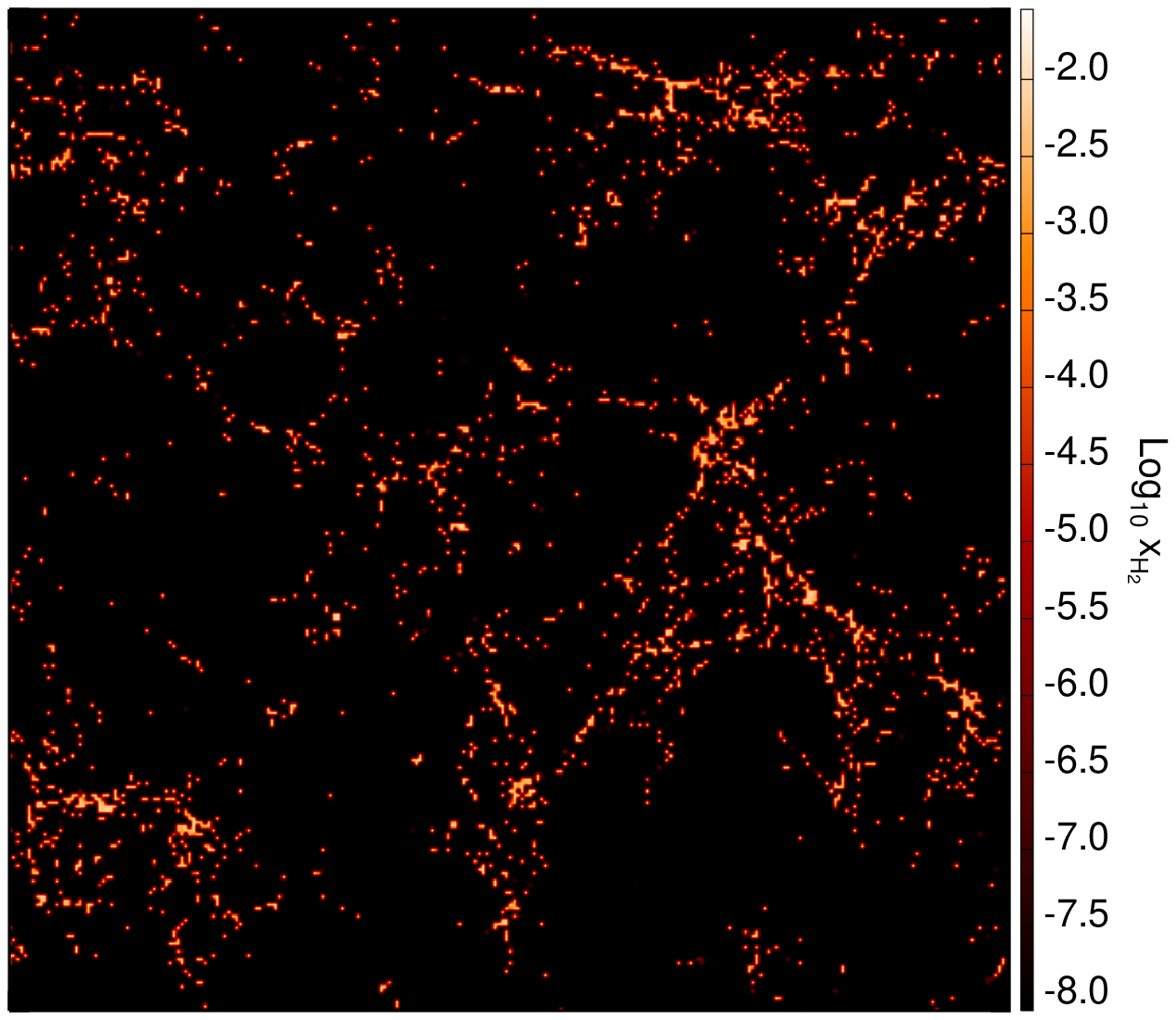}
\vspace{-0.25cm}
\caption{ \small 
Simulation maps in log$_{10}$ scale of the gas entropy in cgs units (left), HI (center) and H$_2$ (right) fractions, as derived by projecting a thin slice of 1/10th the box length along the third direction. 
}
\label{fig:maps}
\end{figure*}
A pictorial view of the simulations performed can be seen in figure~\ref{fig:maps}, where maps of gas entropy, HI and H$_2$ fractions are shown at $ z=4.9$, after the HM UV background sets in. It is easy to recognise the filamentary shielded structures in the cosmic web as the {\it loci} where HI gas is found and H$_2$ molecules form.
The knots represent cold collapsing regions where gas is driven by non-equilibrium cooling and star formation takes place.
\\
From a quantitative point of view, given HI and H$_2$ masses, cold gas is often quantified in terms of the comoving neutral and H$_2$ mass density parameters, $\Omega_{\rm neutral}$ and $\Omega_{\rm H_2}$, defined as the corresponding mass densities divided by $\rho_{0,crit} $.
Observational determinations (see e.g. \cite{PH2020} and references therein) suggest that neutral gas at $z\lesssim 5$ features values around $\Omega_{\rm neutral} \sim 10^{-3} $, increasing smoothly towards the epoch of reionization.
On the contrary, $\Omega_{\rm H_2}$ observations in the (sub-)mm and IR \cite{PH2020, Riechers2020, Garratt2021} give a more complex picture of molecular gas, with peak values of $\Omega_{\rm H_2} \sim 10^{-4} $ at $z\sim 2-4$ and a drop of at least one dex at earlier times.
In this respect, impressive progresses have been made by recent ALMA data in constraining $\rm H_2$ gas up to $z\sim 7$ and in raising previously derived indirect lower limits at $z\simeq 2$-3 \cite{Berta2013}.
\begin{figure}
\centering
\includegraphics[width=0.35\textwidth, trim=0 18 0 30, clip]{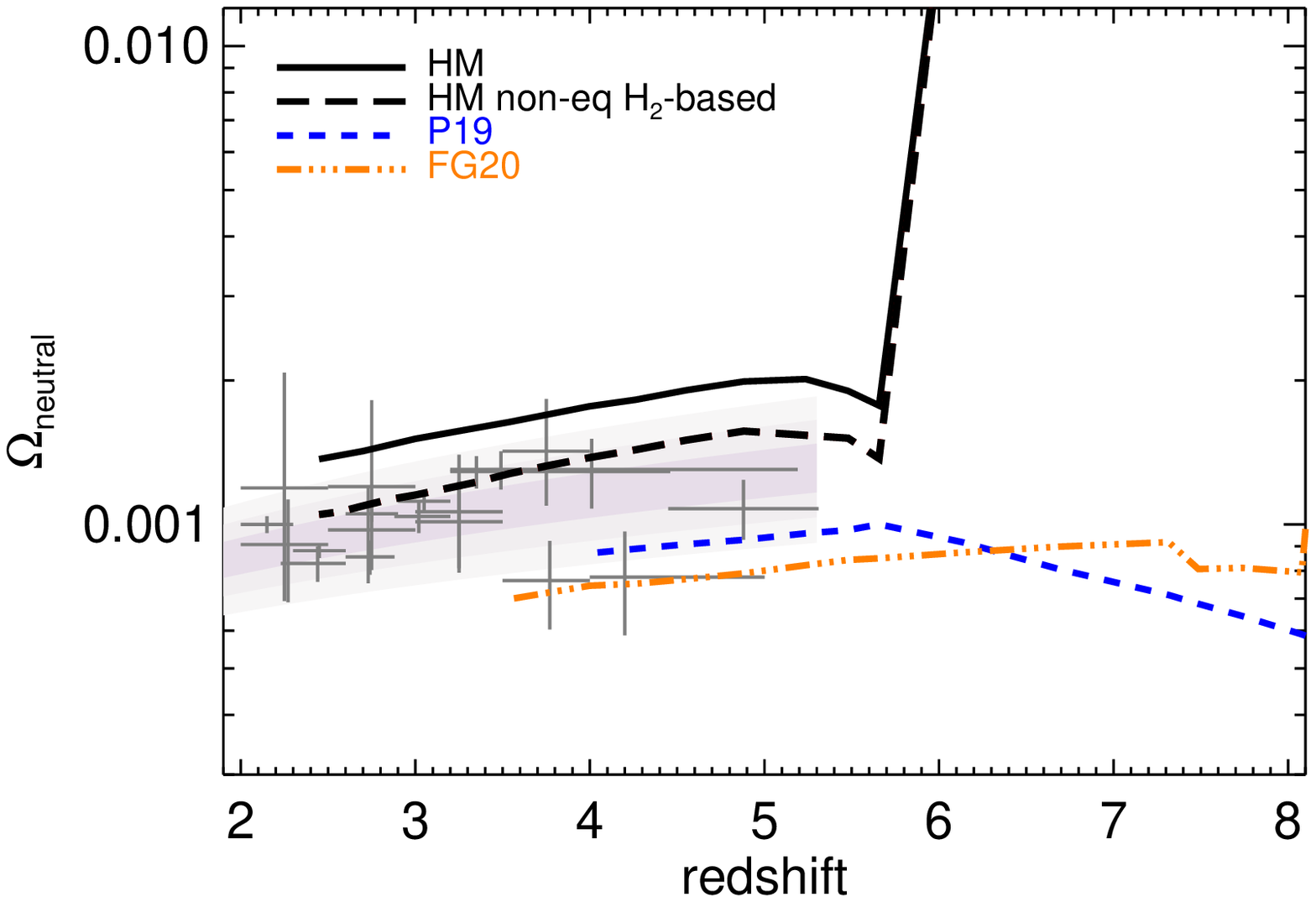}
\hspace{-0.5cm}
\includegraphics[width=0.35\textwidth, trim=0 18 0 30, clip]{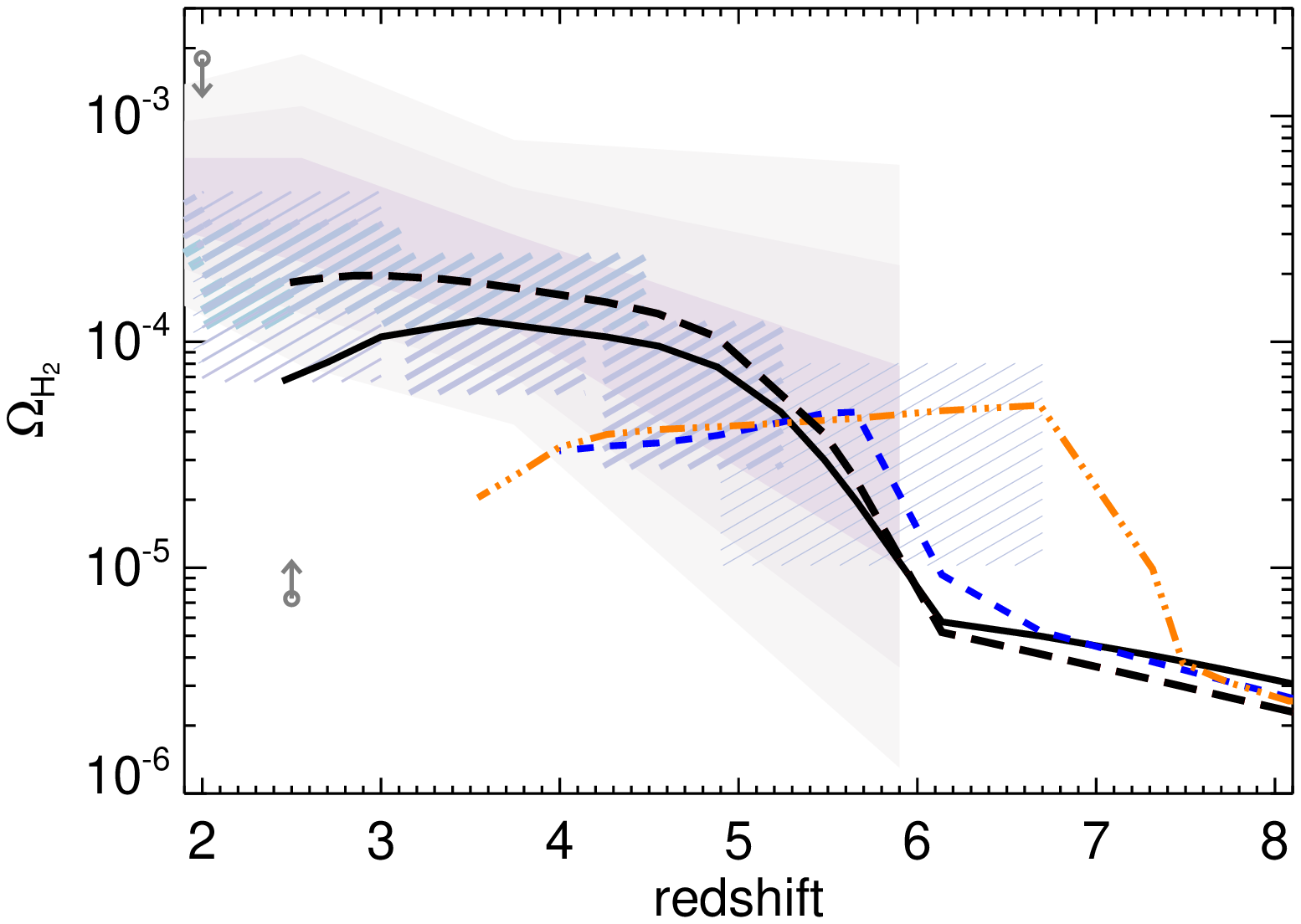}
\hspace{-0.5cm}
\includegraphics[width=0.35\textwidth, trim=0 18 0 25, clip]{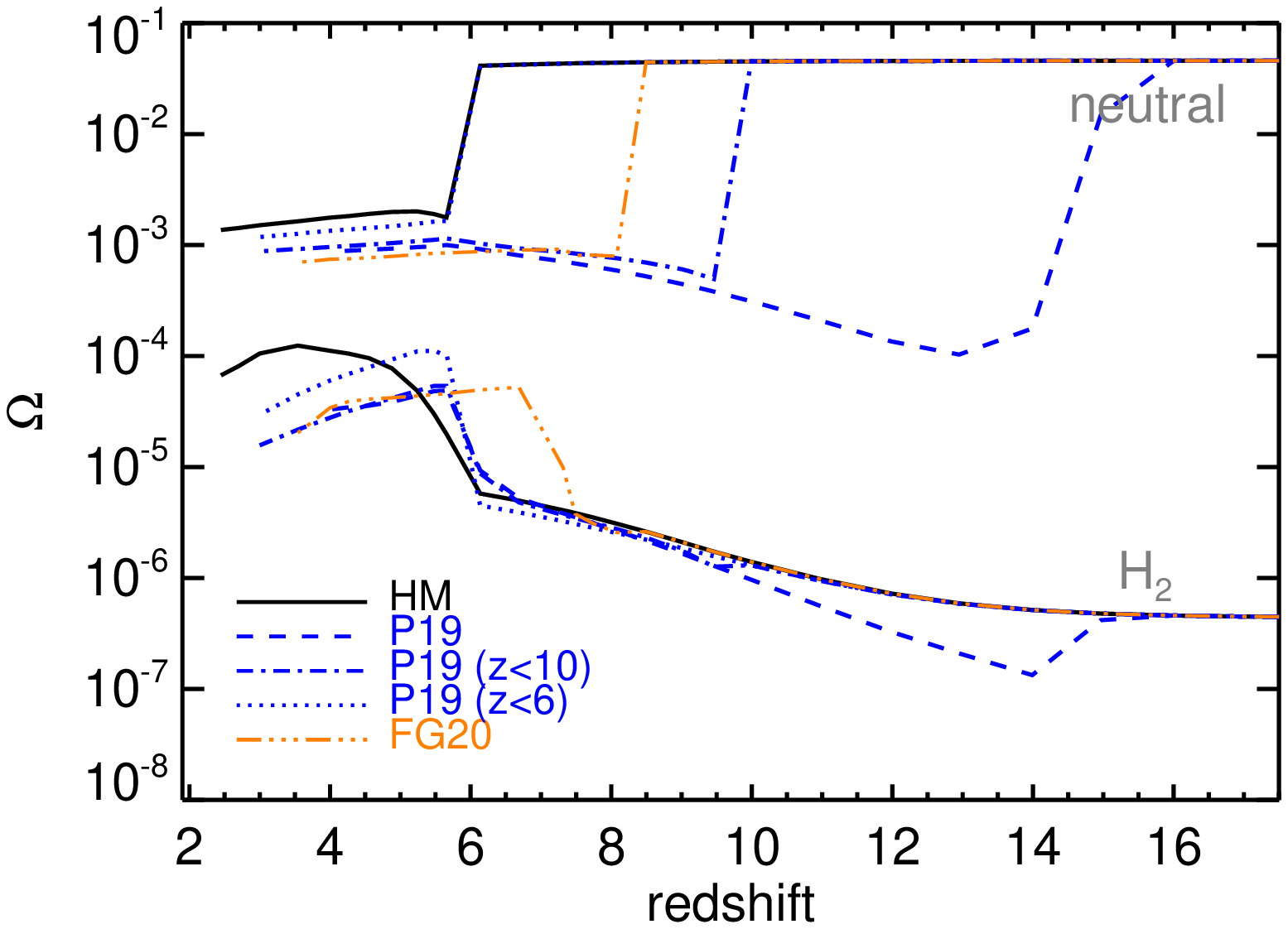}
\vspace{-0.7cm}
\caption{ \small
 $ \Omega_{\rm neutral} $ (left) and $\Omega_{\rm H_2}$ (center) redshift evolution from simulations including HM (black solid lines), P19 (blue short-dashed lines) and FG20 (orange dot-dot-dot-dashed lines) UV backgrounds.
For the HM case, results from the simulation run including non-equilibrium H$_2$-density dependent star formation (non-equilibrium H$_2$-based star formation) are shown, as well (black long-dashed lines).
Data points and shaded areas correspond to recent observational determinations \cite{PH2020, Riechers2020, Decarli2020, Garratt2021}, while $z<3$  lower and upper limits for $\Omega_{\rm H_2}$ are the estimates by \cite{Berta2013} and \cite{Klitsch2019} , respectively.
On the right, both mass density parameters are shown on a larger redshift range. Two additional cases adopting non-null P19 UV rates at $z<10$ (blue dot-dashed lines) and $z<6$ (blue dotted lines), respectively, are also displayed.
}
\label{fig:omega}
\end{figure}
Figure~\ref{fig:omega} displays the trends expected for $ \Omega_{\rm neutral} $ and $\Omega_{\rm H_2}$ from simulations including HM, P19 and FG20 UV backgrounds.
They all include HI and H$_2$ self-shielding, the various H$_2$ formation channels, as well as density-dependent star formation and feedback effects.
In the HM case, we also check the implications of H$_2$-density dependent star formation based on the non-equilibrium species abundance calculations (non-equilibrium H$_2$-based star formation).
In the left and central panels, model predictions are compared to our collection of recent HI and H$_2$ observational determinations.
From the numerical results, it is clear that the various UV background models are roughly consistent with $\Omega _{\rm neutral} $ observational data.
The HM non-equilibrium H$_2$-based case leads to slightly lower $\Omega _{\rm neutral} $ values, in better agreement with observations.
This is a consequence of the smaller star formation feedback and the consequently larger amounts of neutral H mass converted into H$_2$.
The behaviour of the $\Omega_{\rm H_2}$ evolution is more strongly affected by modelling assumptions, although it is evident that P19 and FG20 backgrounds are in tension with the trend inferred by observations.
The HM non-equilibrium H$_2$-based case features $\Omega_{\rm H_2}$ values that are larger by up to a factor of a few with respect to the HM scenario.
Consistently with what just said about $ \Omega _{\rm neutral} $, this is a result of the less efficient star formation feedback in destroying molecules.
We warn the reader that all the several physical and chemical processes involved here play a role in shaping HI and H$_2$ gas.
UV radiation and local chemical heating or cooling cumulate to provide the desired neutral and ionised H fractions.
Molecule formation paths are usually led by the H$^-$ channel in pristine gas and by dust grain catalysis in enriched one.
Gas self-shielding is crucial to obtain reasonable abundances at all $z$, instead, the individual effects of photoelectric or cosmic-ray heating induce only small variations on mass density parameters.
These channels, that are effective in quite heterogeneous conditions, can explain the observationally inferred lack of environmental dependence in molecular-mass build-up at $z \lesssim 3.5$ \cite{Darvish2018} and the large molecular fractions of 50\% or higher, recently inferred at $z\sim 4$-6 \cite{Tacconi2020, DZ2020}, can be easily justified by local H$_2$ formation both in pristine and in enriched cold material.
The depletion times implied by the H$_2$ gas mass densities result of the order of 1/10th the Hubble time at all $z$.
This is a very interesting piece of information, since it suggests that cosmic gas can collapse and structures can form even in the early epoch of reionization.
We note that the sharp increase for both HM models at $z > 6$ is due to the UV rates proposed by \cite{HM1996}. Since they are equal to zero at those early cosmic epochs, early cosmic gas is not photoheated above $10^4\,\rm K$ and remains mostly neutral with $\Omega _{\rm neutral} $ reaching the baryon density parameter at $z\simeq 6$, as neatly visible in the right panel.
In the P19 and FG20 cases the UV background sets on at  $z\simeq 15$ and $z\simeq 8$, respectively, and, indeed, an analogous convergence is found at those redshifts.
Of course, the effects of the UV background onset depend on the strength of the rates at different $z$.
This is confirmed by the two alternative P19 scenarios displayed, where the UV background is assumed to set on at $z=10$ or $z=6$ with null rates at higher $z$.
As said, H$_2$ evolution rely on the available charges in the cosmic medium.
At high $z$, molecular fractions are always low, close to typical early cosmic values of $\sim 10^{-6}$.
While cosmic evolution progresses, the injection of UV radiation can either enhance $\Omega_{\rm H_2}$ through ionization of neutral particles (as is the case for HM at $z<6$) or inhibit it when UV rates are too large (as it happens for P19 and FG20 at $z<8$, in which regime, despite partially boosted, $\Omega_{\rm H_2}$ hardly reaches values of $10^{-4}$ due to dominant H$_2$ destruction).
In fact, by a direct comparison of the different UV rates, one can see that HM photoionization rates are much smaller than P19 ones at $z\lesssim 6$ and FG20 ones at $z\lesssim 8$.
The typically larger P19 and FG20 UV rates explain why corresponding $\Omega _{\rm neutral} $ values, albeit compatible with observational limits, are slightly lower than HM ones at $ z < 6 $.
The high-$z$ tail of the P19 model is additionally responsible for possible H ionization and H$_2$ destruction at $z \gtrsim 10$.

\section{Conclusions}
\label{sec:conclusions}
We have quantified the evolution of cold atomic and molecular cosmic gas and interpreted state-of-the-art observations in the (sub-)mm and IR ranges by means of three-dimensional cosmological N-body hydrodynamic chemistry simulations following: time-dependent non-equilibrium abundance calculations, the impact of different UV backgrounds, HI and H$_2$ self-shielding, dust grain catalysis, photoelectric effect, cosmic-ray heating, star formation, stellar feedback and production of heavy-elements from stars with different masses (SNII, AGB, SNIa) and metallicities.
This is among the first studies addressing cosmic HI and H$_2$ evolution by three-dimensional cosmological hydrodynamic simulations that include non-equilibrium chemistry networks and such levels of detail in the implementation of cosmic-gas processes.
We find that:\\
-- the evolution of cold cosmic gas ($\Omega_{\rm neutral}$ and $\Omega_{\rm H_2}$) obtained with basic time-dependent non-equilibrium chemistry is broadly consistent with the latest observational determinations;\\
-- details about the UV background are modest for $\Omega_{\rm neutral}$ (within the $z$ range where observations are available), but can be relevant for $\Omega_{\rm H_2}$, in particular early ($z \gtrsim 8$) UV onsets could require either lower UV rates or larger gas self-shielding;\\
-- a non-equilibrium H$_2$-based star formation prescription coupled to the UV background leads to results that are in slightly better agreement with HI and H$_2$ determinations;\\
-- H$_2$ evolution is globally driven by the H$^-$ channel, with a relevant contribution from H$_2$ grain catalysis at high metallicities and on local scales: these essential paths can explain the observationally inferred lack of environmental dependence in molecular-mass build-up at $z \lesssim 3.5$ and the large H$_2$ fractions detected at $z \simeq 4$-6;\\
-- a precise determination of dust grain temperature has little relevance and is generally subdominant to metallicity effects;\\
-- cosmic-ray heating in star forming regions, quantified in accordance to different models for the cosmic-ray ionization rate, affects the H$_2$ abundance at ten-per-cent level, globally;\\
-- wind feedback impact results mostly at lower redshift, while SN explosions affect early star forming regimes, with effects from IMF variations remaining usually small at all times;\\
-- despite the insurgence of UV radiation, H$_2$ depletion times reach values lower than the Hubble time within the first half Gyr.\\
These results stress the relevance of non-equilibrium chemistry treatments for reproducing the cosmic abundances of atomic and molecular species.
They also suggest the need to perform dedicated numerical implementations to understand in depth the physics of cosmic gas and to improve upon the works largely available in the literature.
We finish by noting that the ability to form significant amounts of H$_2$ and to reach short depletion times during the epoch of reionization means that future discoveries of molecular-rich star forming galaxies at early times will be possible in the next years.
Therefore, new data from upcoming international facilities will be decisive to shed light on the still unanswered questions about the origin of chemical species and the whole cosmic baryon cycle.

\bibliography{template.bib}

\begin{thebibliography}{26}

\bibitem{Springel2005}
V.~{Springel}, \mnras \textbf{364}, 1105 (2005), \texttt{0505010}

\bibitem{Maio2007}
U.~{Maio}, K.~{Dolag}, B.~{Ciardi}, L.~{Tornatore}, \mnras \textbf{379}, 963
  (2007), \texttt{0704.2182}

\bibitem{Maio2010}
U.~{Maio}, B.~{Ciardi}, K.~{Dolag}, L.~{Tornatore}, S.~{Khochfar}, \mnras
  \textbf{407}, 1003 (2010), \texttt{1003.4992}

\bibitem{Maio2011}
U.~{Maio}, S.~{Khochfar}, J.L. {Johnson}, B.~{Ciardi}, \mnras \textbf{414},
  1145 (2011), \texttt{1011.3999}

\bibitem{Maio2015}
U.~{Maio}, E.~{Tescari}, \mnras \textbf{453}, 3798 (2015), \texttt{1509.03637}

\bibitem{Rahmati2013}
A.~{Rahmati}, A.H. {Pawlik}, M.~{Rai{\v{c}}evi{\'c}}, J.~{Schaye}, \mnras
  \textbf{430}, 2427 (2013), \texttt{1210.7808}

\bibitem{DB1996}
B.T. {Draine}, F.~{Bertoldi}, \apj \textbf{468}, 269 (1996),
  \texttt{astro-ph/9603032}

\bibitem{HM1996}
F.~{Haardt}, P.~{Madau}, \apj \textbf{461}, 20 (1996),
  \texttt{astro-ph/9509093}

\bibitem{P2019}
E.~{Puchwein}, F.~{Haardt}, M.G. {Haehnelt}, P.~{Madau}, \mnras \textbf{485},
  47 (2019), \texttt{1801.04931}

\bibitem{FG2020}
C.A. {Faucher-Gigu{\`e}re}, \mnras \textbf{493}, 1614 (2020),
  \texttt{1903.08657}

\bibitem{Tornatore2007}
L.~{Tornatore}, S.~{Borgani}, K.~{Dolag}, F.~{Matteucci}, \mnras \textbf{382},
  1050 (2007), \texttt{0705.1921}

\bibitem{HM1979}
D.~{Hollenbach}, C.F. {McKee}, \apjs \textbf{41}, 555 (1979)

\bibitem{BT1994}
E.L.O. {Bakes}, A.G.G.M. {Tielens}, \apj \textbf{427}, 822 (1994)

\bibitem{Cazaux2004}
S.~{Cazaux}, A.G.G.M. {Tielens}, \apj \textbf{604}, 222 (2004)

\bibitem{Omukai2005}
K.~{Omukai}, T.~{Tsuribe}, R.~{Schneider}, A.~{Ferrara}, \apj \textbf{626}, 627
  (2005)

\bibitem{DL1984}
B.T. {Draine}, H.M. {Lee}, \apj \textbf{285}, 89 (1984)

\bibitem{Padovani2018}
M.~{Padovani}, A.V. {Ivlev}, D.~{Galli}, P.~{Caselli}, \aap \textbf{614}, A111
  (2018), \texttt{1803.09348}

\bibitem{PH2020}
C.~{P{\'e}roux}, J.C. {Howk}, \araa \textbf{58}, 363 (2020),
  \texttt{2011.01935}

\bibitem{Riechers2020}
D.A. {Riechers}, L.A. {Boogaard}, R.~{Decarli}, J.~{Gonz{\'a}lez-L{\'o}pez},
  I.~{Smail}, F.~{Walter}, M.~{Aravena}, C.L. {Carilli}, P.C. {Cortes},
  P.~{Cox} et~al., \apjl \textbf{896}, L21 (2020), \texttt{2005.09653}

\bibitem{Garratt2021}
T.K. {Garratt}, K.E.K. {Coppin}, J.E. {Geach}, O.~{Almaini}, W.G. {Hartley},
  D.T. {Maltby}, C.J. {Simpson}, A.~{Wilkinson}, C.J. {Conselice}, M.~{Franco}
  et~al., \apj \textbf{912}, 62 (2021), \texttt{2103.08613}

\bibitem{Berta2013}
S.~{Berta}, D.~{Lutz}, R.~{Nordon}, R.~{Genzel}, B.~{Magnelli}, P.~{Popesso},
  D.~{Rosario}, A.~{Saintonge}, S.~{Wuyts}, L.J. {Tacconi}, \aap \textbf{555},
  L8 (2013), \texttt{1304.7771}

\bibitem{Decarli2020}
R.~{Decarli}, M.~{Aravena}, L.~{Boogaard}, C.~{Carilli},
  J.~{Gonz{\'a}lez-L{\'o}pez}, F.~{Walter}, P.C. {Cortes}, P.~{Cox}, E.~{da
  Cunha}, E.~{Daddi} et~al., \apj \textbf{902}, 110 (2020), \texttt{2009.10744}

\bibitem{Klitsch2019}
A.~{Klitsch}, C.~{P{\'e}roux}, M.A. {Zwaan}, I.~{Smail}, D.~{Nelson},
  G.~{Popping}, C.C. {Chen}, B.~{Diemer}, R.J. {Ivison}, J.R. {Allison} et~al.,
  \mnras \textbf{490}, 1220 (2019), \texttt{1909.08624}

\bibitem{Darvish2018}
B.~{Darvish}, N.Z. {Scoville}, C.~{Martin}, B.~{Mobasher}, T.~{Diaz-Santos},
  L.~{Shen}, \apj \textbf{860}, 111 (2018), \texttt{1805.10291}

\bibitem{Tacconi2020}
L.J. {Tacconi}, R.~{Genzel}, A.~{Sternberg}, \araa \textbf{58}, 157 (2020),
  \texttt{2003.06245}

\bibitem{DZ2020}
M.~{Dessauges-Zavadsky}, M.~{Ginolfi}, F.~{Pozzi}, M.~{B{\'e}thermin}, O.~{Le
  F{\`e}vre}, S.~{Fujimoto}, J.D. {Silverman}, G.C. {Jones}, L.~{Vallini},
  D.~{Schaerer} et~al., \aap \textbf{643}, A5 (2020), \texttt{2004.10771}

\end{thebibliography}

\end{document}